\pdfoutput=1

\documentclass{IEEEtran4PSCC}

\usepackage{ifpdf}

\usepackage{cite}

\ifCLASSINFOpdf
\else
\fi
\usepackage[cmex10]{amsmath}
\usepackage[ruled,vlined]{algorithm2e}
\usepackage{units} 

\usepackage{amssymb}
\usepackage{xcolor}
\usepackage{graphicx}

\usepackage{booktabs}
\usepackage{hyperref}
\definecolor{mydarkblue}{rgb}{0,0.08,0.45}
\hypersetup{
    colorlinks=true,
    linkcolor={mydarkblue},
    filecolor=magenta,      
    urlcolor=black, 
    citecolor={mydarkblue},
}
\usepackage{amsmath,amsfonts,bm}

\newcommand{\explainEq}[1]{\overset{\underset{\mathrm{(#1)}}{}}{=}}
\newcommand{\explainLeq}[1]{\overset{\underset{\mathrm{(#1)}}{}}{\leq}}

\def\Figref#1{Figure~\ref{#1}}

\def\secref#1{section~\ref{#1}}

\def\eqref#1{equation~\ref{#1}}

\def\plaineqref#1{\ref{#1}}
\def\plaineqrefp#1{(\ref{#1})}
\def\eqrefp#1{equation~(\ref{#1})} 

\def\Algref#1{Algorithm~\ref{#1}}

\def\transpose{{\intercal}}

\def\ST{\text{s.t.}} 

\def\1{\bm{1}}

\def\ry{{\textnormal{y}}}

\def\rvd{{\mathbf{d}}}

\def\vzero{{\bm{0}}}

\def\vd{{\bm{d}}}

\def\vp{{\bm{p}}}

\def\vv{{\bm{v}}}

\def\vx{{\bm{x}}}
\def\vy{{\bm{y}}}

\def\mA{{\bm{A}}}

\def\mD{{\bm{D}}}

\def\mG{{\bm{G}}}

\def\mI{{\bm{I}}}

\def\mQ{{\bm{Q}}}

\def\mV{{\bm{V}}}

\DeclareMathAlphabet{\mathsfit}{\encodingdefault}{\sfdefault}{m}{sl}
\SetMathAlphabet{\mathsfit}{bold}{\encodingdefault}{\sfdefault}{bx}{n}

\def\gD{{\mathcal{D}}}

\def\gL{{\mathcal{L}}}

\def\gN{{\mathcal{N}}}

\def\gX{{\mathcal{X}}}
\def\gY{{\mathcal{Y}}}

\def\sR{{\mathbb{R}}}

\newcommand{\E}{\mathbb{E}}

\newcommand{\R}{\mathbb{R}}

\DeclareMathOperator*{\argmin}{arg\,min}

\DeclareMathOperator{\Tr}{Tr}

\begin{document}
%


\title{Energy Resource Control via Privacy Preserving Data}


\author{\IEEEauthorblockN{Xiao Chen\IEEEauthorrefmark{2},
Thomas Navidi\IEEEauthorrefmark{3},
Ram Rajagopal\IEEEauthorrefmark{2}\IEEEauthorrefmark{3} 
}
\IEEEauthorblockA{\IEEEauthorrefmark{2} Civil and Environmental Engineering, Stanford University}
\IEEEauthorblockA{\IEEEauthorrefmark{3} Electrical Engineering, Stanford University\\
\{markcx, tnavidi, ramr\}@stanford.edu}

}

\maketitle 

\begin{abstract}
Although the frequent monitoring of smart meters enables granular control over energy resources, it also increases the risk of leakage of private information such as income, home occupancy, and power consumption behavior that can be inferred from the data by an adversary. We propose a method of releasing modified smart meter data so specific private attributes are obscured while the utility of the data for use in an energy resource controller is preserved. The method achieves privatization by injecting noise conditional on the private attribute through a linear filter learned via a minimax optimization. The optimization contains the loss function of a classifier for the private attribute, which we maximize, and the energy resource controller's objective formulated as a canonical form optimization, which we minimize. We perform our experiment on a dataset of household consumption with solar generation and another from the Commission for Energy Regulation that contains household smart meter data with sensitive attributes such as income and home occupancy. We demonstrate that our method is able to significantly reduce the ability of an adversary to classify the private attribute while maintaining a similar objective value for an energy storage controller.
\end{abstract}

%






\section{Introduction}
Traditionally, the power grid has been managed by the producers and grid operators with information primarily exchanged among the large asset owners with little feedback from its end users. However, the push for renewable energy sources has brought about the rise of distributed energy resources (DERs) that lie under the control of many smaller and disparate users causing a paradigm shift in the flow of information in the grid. The successful operation of DERs and other smart grid technologies depends on the exchange of large amounts of data from many different end users~\cite{BilevelBernstein, ZhangCoop,NavidiCdc}. Unfortunately, it may be unrealistic to assume the data will be available without consideration of the privacy concerns the data owners may face. It has been demonstrated that the increased granularity of data required for smart grid operation enables the inference of personal information~\cite{inferDr}, which suggests data owners may be reluctant to exchange their data without some effort towards privacy preservation.
%
  


Studies have investigated various approaches to protect smart meter data privacy using a number of different metrics. While detailed surveys are given in \cite{jawurek2012sok,komninos2014survey}, we briefly cover a few popular solutions here. Aggregating data or its statistics has been considered \cite{buescher2017two, corrigan2017prio} to provide user privacy since the aggregated data does not reflect any specific meter data above a certain aggregation size. Another approach at privatization comes from differential privacy \cite{DworkC2008Survey}, which is widely adopted in privacy mechanism design and analysis in the context of energy data \cite{sankar2012smart, han2016event, chin2017privacy, eibl2017differential, ZhouAndersonLow2019ACC}. Specifically, studies \cite{sankar2012smart, han2016event} and \cite{chin2017privacy} proposed several frameworks for reducing the mutual information between raw data and privatized data (e.g. power profiles). Approaches proposed in \cite{eibl2017differential}  investigated the differential privacy effect with some noise injection (e.g. Laplace noise). It showed the aggregation group size must be of the order of thousands of smart meters in order to have reasonable utility. And \cite{ZhouAndersonLow2019ACC} explored how much noise is required to be added to the data in order to achieve a certain level of differential privacy for existing Laplace mechanism in the context of solving optimal power flow.  


We distinguish our studies by focusing on developing a methodology that learns an optimal noise injection for balancing the trade off between privacy and data utility, thus, preserving as much utility in the data as possible. It differs from strict differential privacy because we use a general notion of privacy that is the reduced correlation between private attributes and the data. This general notion of privacy gives us the flexibility to maintain the utility of the data while still eliminating an adversary's ability to recognize certain private attributes. Since many applications of smart meter data involve their use in optimization procedures, we define the utility as the performance achieved when such data is used for optimal control. We consider a scenario where individual owners of DERs, such as battery storage systems, wish to privatize their data before releasing it to a DER aggregator to make optimal control decisions on their behalf, which can have applications in the context of~\cite{BilevelBernstein, ZhangCoop}.

Our primary contributions are a minimax approach to generate realistic meter data that is decorrelated from sensitive attributes while maintaining limited performance loss of a cost minimization optimal control algorithm using battery storage. Additionally, we developed a parallelized method that can be easily incorporated in modern deep learning architectures. The correlation of data privatized by our method with sensitive attributes and the performance of a control algorithm is evaluated on two real datasets of residential power demand: one with synthetic sensitive labels and one with real labels. We demonstrate that our method is able to decrease the classification accuracy of an adversary by over 20\% while maintaining the performance of the optimization to within 10\% over both datasets.

The rest of the paper is organized as follows: we describe the energy resource control in \secref{sec:der:ctrl}, control with privatized data generated from the minimax learning algorithm in \secref{sec:ctrl:priv:demand}, experiments and results on the two datasets in \secref{sec:experiments:results}, and the Conclusion in Section \secref{sec:conclusion}.

\section{Energy Resource Control}
\label{sec:der:ctrl}
%
\subsection{Notation}
We use bold letters for vectors and matrices and regular letters for scalars. Given two vectors $\vx$ and $\vy \in \sR^n$, $\vx \geq \vy$ represents the element-wise order $\vx(i) \geq \vy(i)$ for $i \in [n]$ where $[n]$ denotes the set $[n]=\{1,\hdots, n\}$. And $\vx \geq 0$ means all elements in the vector are not less than the scalar zero. We make the dependence on the underlying probability distribution $P$ when we write
expectations (e.g. $\E_P [X]$ where $X$ denotes a random variable). The Frobenius norm of a matrix $\mA$ is $||\mA||_F$. We write $\nabla_{\theta} \gL(\theta; X)$ or $\mathsf{d}\gL(\theta; X)$, where we typically mean differentiation of the loss function $\gL$ with respect to the parameter $\theta \in \sR^n$. $\gN$ stands for Normal (or Gaussian) distribution and $\sR_+$ denotes the non-negative real numbers.  We use $:=$ to represent "define as." All the vectors are column vectors by default unless we explicitly address otherwise in a specific context. 

\subsection{Battery storage control}

\textbf{\emph{Control with deterministic demand}}: Consider a basic battery control problem with the goal of minimizing the energy cost given a prescribed price $\vp \in \sR^{H}$ where $H$ is the time horizon that is typically $24$ if it is an hourly price. An uncontrollable electricity demand is specified as $\vd \in \sR_{+}^{H}$. We denote the decision variables for battery control to be $\vx$ and expand it into $\vx_{in}, \vx_{out}, \vx_{s} \in \sR_{+}^{H}$ that represents the charging, discharging, and the amount of charge in storage, i.e. $\vx^{\transpose} = [\vx_{in}^{\transpose}, \vx_{out}^{\transpose}, \vx_{s}^{\transpose}] $. The battery optimal control is formulated as follows (\textbf{Problem\plaineqref{ctrl:privD:prob1}}):
\begin{subequations}
\begin{small}\label{ctrl:privD:prob1}
\begin{align}
\begin{split}
 \min_{\vx} & \vp^{\transpose}(\vx_{in} - \vx_{out} + \vd)_{+} + \beta_1||\vx_{in}||_2^2 + \beta_2||\vx_{out}||_2^2 \\
 & + \beta_3||\vx_s - \alpha B||_2^2
 \end{split} \label{ctrl:privD:prob1:obj} \\
 \ST\quad & \vx_{s}(j+1) = \vx_{s}(j) - \frac{1}{\eta_{out}} \vx_{out}(j) + \eta_{in} \vx_{in}(j) \quad \forall j \in [H] \label{ctrl:privD:prob1:c1}\\
 & \vx_{s}(1) = {B}_{init} \label{ctrl:privD:prob1:c2}\\
 & 0 \leq \vx_{in} \leq c_{in} \label{ctrl:privD:prob1:c3} \\
 & 0 \leq \vx_{out} \leq c_{out} \label{ctrl:privD:prob1:c4} \\
 & 0 \leq \vx_{s} \leq B.  \label{ctrl:privD:prob1:c5}%
\end{align}  
\end{small}%
\end{subequations}%
The linear term (with respect to $\vx$) in the  objective is the cost of electricity when there is no value for selling the energy back to the grid. This represents a situation where there are no net-metering incentives. The quadratic penalty terms $\beta_1||\vx_{in}||_2^2$ and $\beta_2||\vx_{out}||_2^2$ are added to
protect the battery state of health in the horizon \cite{liu2018customer}. The term $\beta_3||\vx_{s} - \alpha B||_2^2$ is added to set the battery state to be close to the target value $\alpha B$ with $B$ as the battery size and $\alpha \in (0, 1)$. $\beta_1, \beta_2, \beta_3$ are hyper-parameters to control these penalties. $c_{in}$ and $c_{out}$ are the charging-in and discharging-out power capacities. And the parameter $\eta_{in}$ and $\eta_{out}$ denote the charging and discharging efficiency (between 0 and 1). The constraint \plaineqrefp{ctrl:privD:prob1:c1} indicates that the battery state in the next timestep equals the current battery state adding up the net charging amount (summing up charging and discharging together). Constraint \plaineqrefp{ctrl:privD:prob1:c2} sets the initial state of the battery to have $B_{init}$. To simplify the notation, we define a set $\gX := \{ \vx | \text{\plaineqrefp{ctrl:privD:prob1:c1} - \plaineqrefp{ctrl:privD:prob1:c5}  are feasible for some } \vx \in \sR^{3H} \}$.
Hence, we use $\vx \in \gX$ to succinctly express that $\vx$ satisfies the battery constraints. We convert the problem \plaineqrefp{ctrl:privD:prob1} into canonical convex form in Appendix \ref{appx:qp:formulate} and develop a paralleled algorithm making use of automatic differentiation, open-source convex solvers, and pytorch\cite{paszke2017automatic}--a popular deep learning framework.

\textbf{\emph{Control with stochastic demand}}: When determining the control with an uncertain demand, we minimize the expected cost under some demand distribution $P$. The objective is slightly changed as follows (\textbf{Problem\ref{ctrl:privD:prob2}}):
\begin{subequations}
\begin{small}\label{ctrl:privD:prob2}
\begin{align}
\begin{split}
\min\gL_u(\vx, \vd) & := \min_{\vx}  \E_{{\vd} \sim P}\big[\vp^{\transpose}(\vx_{in}-\vx_{out}+{\vd})_{+}\big] \\
&+ \beta_1||\vx_{in}||_2^2 +\beta_2||\vx_{out}||_2^2  + \beta_3||\vx_s - \alpha B||_2^2 \label{ctrl:privD:prob2:obj}%
\end{split}  \\
\ST \quad & \vx \in \gX.
\end{align}
\end{small}
\end{subequations}
%
%
\section{Control with Privatized Demand}
\label{sec:ctrl:priv:demand}%
Protecting privacy in our context means reducing the correlation between the smart meter data and the sensitive attribute of the data owner, e.g. income or square-footage of the house. We justify why such a consideration of privacy protection is useful in practice in \secref{ctrl:privD:sec:privdata}

\subsection{Revealing privacy from data} \label{ctrl:privD:sec:privdata}%
In this section, we consider a simple scenario that the sensitive information is a binary label, such as a small or large home, which can be inferred from smart meter data. Given the raw demand $\rvd \in \sR_{+}^{H}$ and sensitive label $\ry \in \{0, 1\}$, the adversary builds a classifier $f_{\psi}$ that takes in demand $\vd$ to estimate $\ry$ with a prescribed loss function $\gL_a$. Specifically, we assume the adversary minimizes the classification loss 
\begin{equation*}
    \min_{\psi} \gL_a\big(f_{\psi}(\vd), \ry\big)
\end{equation*} 
to infer the private information $\ry$. A popular choice of classification loss is cross-entropy loss (or log-loss) \cite{Lin_2017_ICCV}%
\begin{equation*}
\begin{split}
    & \min_{\psi}\Big\{- \ry\log(f_{\psi}(\vd))-(1-\ry)\log\big(1-f_{\psi}(\vd)\big)\Big\} 
\end{split}
\end{equation*}
when $\ry$ is a binary variable. The classifier $f_{\psi}$ is parameterized by $\psi$ and can be a neural network that outputs an estimate of the probability of the positive label. 
Previous studies \cite{beckel2014revealing, chen2018understanding} showed that estimating a sensitive label such as income or square-footage of the house reaches $69\%$ accuracy using features of smart meter data and models like support vector machine or random forest. We use an alternative neural network model that leverages the daily power consumption (demand) and achieve state of the art accuracy of the private label. More details can be found in \secref{sec:experiments:results}.    

\subsection{Control with private demand}
Our goal is to minimize energy cost incorporating of privacy protection. Specifically, we design a \emph{data generator} that creates a perturbed version of the raw demand data in a way that increases the adversarial classification loss, while enabling an optimal controller to minimize the energy cost. From a modeling perspective, we have a minimax problem (\textbf{Problem\ref{ctrl:privD:prob3}}):
\begin{subequations}\label{ctrl:privD:prob3}
\begin{small}
\begin{align}
& \min_{\mG} \gL_u\big( \tilde{\vx}^*(\tilde{\vd}), \vd\big) - \lambda_a \gL_a(f(\tilde{\vd}), \vy ) \\
\ST & \text{ } \tilde{\vd} = \vd + \mG  \begin{bmatrix}\boldsymbol{\varepsilon} \\ \vy \end{bmatrix}, \boldsymbol{\varepsilon}\sim \gN(0, \mI)\\
& \tilde{\vx}^*(\tilde{\vd}) = \arg\min_{\vx \in \gX} \gL_u(\vx, \tilde{\vd}), \label{ctrl:privD:prob3:3c}%
\end{align}%
\end{small}%
\end{subequations}%
%
%
where the parameter $\mG$ is a matrix that affects the distribution of $\tilde{\vd}$. In this case, we consider a linear transformation of Gaussian noise $\boldsymbol{\varepsilon}$. 
Variable $\vy$ is the one-hot encoding of the sensitive binary label, 
and $f_{\psi}$ is a classifier that takes in the perturbed demand data and predicts the corresponding label private label. 
The $\gL_u$ stands for utility loss. It is important to note that $\gL_u$ in the objective uses the raw demand to evaluate the cost of the control decisions determined using the perturbed demand. This represents the case where the storage unit acts on the perturbed information, but the real world value is based on the original raw data.

In order to solve the non-trivial optimization \plaineqrefp{ctrl:privD:prob3}, we simplify the constraints and make use of adversarial training that is further explained in  \secref{ctrl:privD:sec:sub:minmax:learn}, which is a common technique in studies of generative adversarial networks (GAN) and their applications \cite{goodfellow2014generative, chen2018unsupervised}.

We add a regularization term $\E||\tilde{\vd} - \vd ||_2^2$ in the objective with an additional hyper-parameter $\kappa$,
\begin{small}
\begin{equation}
\min_{\mG} \gL_u\big( \tilde{\vx}^*(\tilde{\vd}), \vd\big) - \lambda_a \gL_a(f(\tilde{\vd}), \vy ) + \kappa \E||\tilde{\vd} - \vd ||_2^2 ,
\end{equation}
\end{small}%
which helps convergence of the training and preserves parts of the demand that are not related to the privacy or utility loss instead of allowing them to be perturbed arbitrarily.

We can denote matrix $\mG = [\Gamma, \mV]$ with $\Gamma \in \R^{H\times H}$ and $\mV \in \R^{H\times 2}$. The altered demand then becomes $\tilde{\vd} = \vd + \Gamma \boldsymbol{\varepsilon} + \mV y$. By denoting $\pi$ to be the prior distribution of one-hot labels, e.g. $\pi = [p, 1-p]^{\transpose}$ where $p$ is the prior probability of a positive label, we can rewrite the distortion regularization as
\begin{small}
\begin{align}
\begin{split}
    & \E(||\tilde{\vd} - \vd ||_2^2) = \E\big[||\vd + \Gamma\boldsymbol{\varepsilon} + \mV \vy - \vd ||_2^2\big] \\
    & = \E\big[ (\Gamma\boldsymbol{\varepsilon}+ \mV \vy)^{\transpose}(\Gamma\boldsymbol{\varepsilon}+ \mV \vy) \big] \\
    & = 
    \E(\boldsymbol{\varepsilon}^{\transpose}\Gamma^{\transpose} \Gamma \boldsymbol{\varepsilon} + \vy^{\transpose}\mV^{\transpose} \mV \vy + \vy^{\transpose}\mV^{\transpose} \Gamma\boldsymbol{\varepsilon} + \boldsymbol{\varepsilon}^{\transpose}\Gamma^{\transpose}\mV\vy) \\
    & \explainEq{i} \E\big[\Tr(\Gamma {\boldsymbol{\varepsilon}\boldsymbol{\varepsilon}^{\transpose}}\Gamma^{\transpose}) + \Tr(\mV \vy \vy^{\transpose}\mV^{\transpose}) \big] \\
    & \explainEq{ii} \Tr(\Gamma \underbrace{\E[ \boldsymbol{\varepsilon}\boldsymbol{\varepsilon}^{\transpose}}_{\mI}]\Gamma^T) \\
    & + \Tr\Big(
    \begin{bmatrix}
    | & | \\
    \vv_1 & \vv_2 \\
    | & | \\
    \end{bmatrix} 
    \underbrace{\begin{bmatrix} 
    p^2 & p(1-p)\\
    p(1-p) & (1-p)^2\\
    \end{bmatrix} }_{\E[\vy\vy^{\transpose}]}
    \begin{bmatrix} 
    - & \vv_1^{\transpose} & - \\
    - & \vv_2^{\transpose} & - \\
    \end{bmatrix}
    \Big) \\
    & \explainEq{iii}  \Tr(\Gamma\Gamma^{\transpose}) + ||p\vv_1 + (1-p)\vv_2||_2^2  \\
    & = ||\Gamma||_{F}^2 + ||\mV \pi||_2^2 \label{ctrl:privD:prob3:reform:c1andc2}
\end{split}    
\end{align} 
\end{small}%
Equality (i) uses the fact that $\boldsymbol{\varepsilon}$ has zero mean. Equality (ii) expands out $\mV$ as column vectors $[\vv_1, \vv_2]$ and expresses $\E[\vy\vy^{\transpose}] = \pi\pi^{\transpose} = \begin{footnotesize}
\begin{bmatrix} p \\ 1-p \end{bmatrix}\begin{bmatrix} p & 1-p\end{bmatrix}\end{footnotesize}$. Rearranging the expressions yields equality (iii).\\ 
Therefore, we can equivalently penalize the Frobenius norm of $\Gamma$ and $l_2$ norm of the vector $\mV\pi$, i.e. $||\Gamma||_{F}^2 + ||\mV\pi||_2^2 $, instead of taking the empirical mean of the demand difference when performing the regularization. To summarize, the data generator determines the filter weight $\mG$ and outputs the perturbed demand $\tilde{\vd}$, while the adversary takes in the altered demand $\tilde{\vd}$ and private labels $\ry$ to try to learn a classifier. 

\subsection{Minimax learning}\label{ctrl:privD:sec:sub:minmax:learn}
We construct two neural networks to perform the roles of the two players, one is for the data generator and the other one is for the adversary. To train the adversary, we minimize the cross-entropy loss $\gL_a$, i.e.  $\min_{\psi}\gL_a\big(f_{\psi}(\tilde{\vd}), \ry\big)$, which follows the loss function mentioned in \secref{ctrl:privD:sec:privdata}. For the generator, we decouple the training into two steps. First, we leverage the loss that is passed from the adversary to update the matrix weight $\mG= [\Gamma, \mV]$, i.e. 
\begin{small}
\begin{align}\label{ctrl:privD:prob4:step1}
\begin{split}
    & (\textbf{step1})  \min_{\mG}-\lambda_a\gL_a\Big( f_{\psi}\big(\vd+\Gamma\boldsymbol{\varepsilon} + \mV \vy\big), \vy \Big) \\
    & + \kappa\big(||\Gamma||_{F}^2 + ||\mV\pi||_2^2  \big) \\
     & \explainEq{i}  \min_{\mG = [\Gamma, \mV]} -\lambda_a\log\Big(1 - f_{\psi}\big(\vd+\Gamma\boldsymbol{\varepsilon} + \mV \vy \big)\Big) \\
     & + \kappa\big(||\Gamma||_{F}^2 + ||\mV\pi||_2^2 \big), 
\end{split}
\end{align}%
\end{small}%
where $\kappa$ is the hyper-parameter that penalizes the distance between $\tilde{d}$ and $d$ implicitly. Equality (i) uses the log-loss as the classification loss for the binary label. The next step is to use the privatized demand $\tilde{\vd}=\vd + \widehat{\mG}\begin{footnotesize}\begin{bmatrix}
 \varepsilon \\ 
 \vy \\
 \end{bmatrix}\end{footnotesize}$ to determine the control by running the following optimization:
\begin{subequations}
\begin{small}\label{ctrl:privD:prob4:step2}
\begin{align}
\begin{split}
(\textbf{step2}) & \argmin_{\vx} \E_{\boldsymbol{\varepsilon}\sim\gN(0, \mI)}\Big\{ \vp^{\transpose}\big(\vx_{in} - \vx_{out} + \tilde{\vd}\big)_{+} \\
& + \beta_1||\vx_{in}||_2^2  + \beta_2||\vx_{out}||_2^2 + \beta_3||\vx_s - \alpha B||_2^2 \Big\}
\end{split}\\
\ST \quad & \vx \in \gX.  
\end{align}%
\end{small}%
\end{subequations}%
The optimal solution of the above convex problem \plaineqrefp{ctrl:privD:prob4:step2} is $\tilde{\vx}^*$, or more specifically $\tilde{\vx}^*(\tilde{\vd})$, because it is a function of the privatized demand, which is aligned with \eqrefp{ctrl:privD:prob3:3c}. The third step calculates the loss, $\gL_u(\tilde{\vx}^{*}, \vd)$, using $\tilde{\vx}^*(\tilde{\vd})$ and the original raw demand expressed as: 
\begin{subequations}
\begin{small}
\begin{align}
\begin{split}
 (\textbf{step3}) & \gL_u(\tilde{\vx}^{*}(\tilde{\vd}), \vd) = \vp^{\transpose}\Big(\tilde{\vx}_{in}^{*}(\tilde{\vd}) - \tilde{\vx}_{out}^{*}(\tilde{\vd}) + \vd \Big)_{+} \\
 & + \beta_1||\tilde{\vx}_{in}^{*}||_2^2  + \beta_2||\tilde{\vx}_{out}^{*}||_2^2 + \beta_3||\tilde{\vx}_s^{*} - \alpha B||_2^2 .
\end{split}
\end{align}\noindent 
\end{small}
\end{subequations}%

We update $\mG$ using gradient descent with the gradient determined by the chain rule. 
Recall that the generator outputs a privatized demand with reduced correlation to the sensitive label that is also used to yield the storage control decisions. Those decisions are evaluated on the cost given the raw demand, thus, the Jacobian of $\mG$ is 
\begin{align}
    g_{\mG} =\nabla_{\mG}{ \gL_u(\tilde{\vx}^{*}, \vd) } = \frac{ \partial \gL_u( \tilde{\vx}^{*}, \vd) }{\partial \vx}\frac{\partial{\vx}}{\partial{\tilde{\vd}}}\frac{\partial{\tilde{\vd}}}{\partial{\mG}}. \label{ctrl:privD:Gamma:backprop} 
\end{align}
In the context of our storage control problem, the first term in \plaineqrefp{ctrl:privD:Gamma:backprop} is 
\begin{small}
\begin{align}
    \frac{ \partial \gL_u({\vx}, \vd) }{\partial \vx} = \begin{cases}
    \mQ {\vx} + \begin{bmatrix}\vp \\
    -\vp \\
    \vzero \end{bmatrix}, \text{ if } \mD {\vx} - \vd > 0\\
    \mQ {\vx}  \quad \text{ otherwise}
    \end{cases},
\end{align}
\end{small}%
where $\mQ$ is given in the Appendix \eqrefp{eq:Q:q:obj:QP:cvx}, 
$\mI$ is the identity matrix, and  $\mD = \begin{bmatrix}\mI & -\mI & \vzero\end{bmatrix}$.


The second term, i.e. $\frac{\partial \vx}{\partial \tilde{\vd}}$, in \plaineqrefp{ctrl:privD:Gamma:backprop} hinges on automatic differentiation through a convex program\cite{amos2017optnet, agrawal2019differentiating}. Because an optimization problem can be viewed as a function mapping the problem data to the primal and dual solutions, we can convert problem \plaineqrefp{ctrl:privD:prob4:step2} to a conic form and calculate the changes of the optimal solution given the perturbations of the problem data.
It leverages the idea of finding a zero solution for the residual map of a homogeneous self-dual embedding derived from the KKT conditions of the convex program\cite{agrawal2019differentiating, ye1994nl, busseti2018solution}.    

The third term in \plaineqrefp{ctrl:privD:Gamma:backprop} is
\begin{small}
\begin{equation}
    \mathsf{d}\mG := \frac{\partial \tilde{\vd}}{\partial \mG} = \begin{bmatrix}
    \frac{\mathsf{d} \tilde{\vd}}{\varepsilon_1} & \hdots & \frac{\mathsf{d} \tilde{\vd}}{\varepsilon_H} &  \frac{\mathsf{d} \tilde{\vd}}{p} & \frac{\mathsf{d} \tilde{\vd}}{1-p} 
    \end{bmatrix} \in \R^{H\times (H+2)},
\end{equation}
\end{small}%
since $\mathsf{d}\tilde{\vd} = \mathsf{d}\mG \begin{bmatrix}
\boldsymbol{\varepsilon} \\
\vy
\end{bmatrix}$.
Thus, all three terms in \eqrefp{ctrl:privD:Gamma:backprop} can be evaluated in the backward pass of the generator training and we can update the filter weight $\mG$ using stochastic gradient decent\cite{bottou2010large}: $\mG_{k+1} := \mG_{k} - \eta_l g_{\mG}$ where $k$ is the iteration step and $\eta_l$ is the learning rate. \\
\textbf{\emph{Remark}}: To summarize, Step 1 shown in \eqrefp{ctrl:privD:prob4:step1} updates the matrix $\mG$ by minimizing the negative classification loss (equivalent to maximizing the classification loss) of the adversary, while maintaining the constraint determined in \plaineqrefp{ctrl:privD:prob3:reform:c1andc2}. Step 2 calculates the optimal control of the storage using the privatized demand. In Step 3, $\mG$ is updated by evaluating the gradient of the energy cost given the control based on the privatized demand. The updates are expressed as
\begin{subequations}
\begin{align}
  &(\text{update1})  \widehat{\mG}_{k+1}  = \mG_k - \eta_{l}^{(k)} \nabla_{G} \gL_a(f_{\psi}(\tilde{\vd}), \vy)\\
  &(\text{update2})  \mG_{k+1}  = \widehat{\mG}_{k+1} - \eta_{l}^{(k)} \nabla_G \gL_u(\tilde{\vx}^{*}, \vd) \\
  & (\text{adversary update}) \psi_{k+1} = \psi_k - \eta_l \nabla_{\psi} \gL_{a}(f_{\psi}(\tilde{\vd}), \vy), 
\end{align}
\end{subequations}
 which run until convergence. We set the learning rates in each step to be equal for simplicity. The training procedure is described in \Algref{Algorithm:minmax:train}. 
\begin{algorithm}[!hbpt]
\KwIn{Demand data $\gD$, label data $\gY$, learning rate $\eta_l$, parameters $\{B, \alpha, \beta_1, \beta_2, \beta_3\}$, and hyper parameters $\kappa_1, \kappa_2$ }
Initialize $\mG_k$, $\psi_k$ at iteration $k=0$ with batch size $m$;\\
 \While{ $\psi$ or $\mG$ has not converged}{
   \nl draw batches of pair $(\vd^{(i)}, \vy^{(i)})$  from demand and label datasets ($\gD, \gY$), $\forall i = 1,\hdots,m$;\\
   \nl Sample batch of Gaussian random vectors $\boldsymbol{\varepsilon}^{(1),\hdots,(m)} \sim \gN(\vzero, \mI)$; \\ 
   \nl $\psi_{k+1} := \psi_k - \eta_l \E [\nabla_{\psi} \gL_a(f_{\psi}(\tilde{\vd}), \vy)]$; \\
   \nl  $\widehat{\mG}_{k+1} := \mG_k - \eta_{l} \E [\nabla_{G} \gL_a(f_{\psi}(\tilde{\vd}), \vy )]$; \\
   \nl  $\mG_{k+1} := \widehat{\mG}_{k+1} - \eta_{l} \E [\nabla_G \gL_u(\tilde{\vx}^{*}, \vd)]$ where $\tilde{\vx}^{*}$ is optimal solution of \plaineqrefp{ctrl:privD:prob4:step2} \\
   (The expected gradient value is approximated as the sample mean of the batch.)
    }
\Return $\mG$ and $\psi$    
\caption{{\bf Minimax learning} \label{Algorithm:minmax:train}}
\end{algorithm}
\subsection{Convergence of the filter }
This subsection focuses on the stability and boundedness of the iterates in our back-propagation that leverage stochastic gradient methods (or some related variants of first-order gradient methods). Using the subgradient property \cite[Chapter~9.1]{boyd2004convex}, $g$ is a subgradient of $f$ at $x$ if 
\begin{align}
    f(y) \geq f(x) + \langle g, y-x \rangle \quad \forall y,  \label{gan:filter:subgrad:lemma}
\end{align}
and assuming $\mG^*$ is a local optimal point; when we apply the step1 and step3 updates $\mG_{k+1} = \mG_k - \eta_l^{(k)}\nabla \gL_a^{(k)} - \eta_l^{(k)}\gL_u^{(k)}$ at the $k$-th iteration, we can obtain the following  relationship 
\begin{subequations}
\begin{footnotesize}
\begin{align}
    & \E[||\mG_{k+1} - \mG^{*}||_2^2] \label{gan:filter:stability:eq1}\\
    = & \E[||\mG_{k} - \eta_{l}^{(k)}(\nabla \gL_a^{(k)} + \nabla \gL_u^{(k)}) - \mG^{*} ||_2^2] \\
    \begin{split}
     = & \E[||\mG_{k} - \mG^{*}||_2^2] - 2\eta_l^{(k)} \E \langle \nabla \gL_a^{(k)} 
    + \nabla \gL_u^{(k)}, \mG_k - \mG^{*} \rangle \\
    & + (\eta_l^{(k)})^2 \underbrace{||\nabla \gL_a^{(k)} + \nabla \gL_u^{(k)}||_2^2}_{\delta_k^2}
    \end{split} \\
    \begin{split}
    & \explainEq{i} \E[||\mG_{k} - \mG^{*}||_2^2] - 2\eta_l^{(k)} \E \langle \nabla \gL_a^{(k)}, \mG_k - \mG^{*} \rangle \\
    & - 2\eta_l^{(k)} \E \langle \nabla \gL_u^{(k)}, \mG_k - \mG^{*} \rangle + (\eta_l^{(k)})^2\delta_k^2
    \end{split} \\
    \begin{split}
    & \explainLeq{ii} \E[||\mG_{k} - \mG^{*}||_2^2] - 2\eta_l^{(k)} \big(\gL_a(\mG_k) - \gL_a^*\big) \\
    & - 2\eta_l^{(k)} \big(\gL_u(\mG_k) - \gL_u^*\big) +  (\eta_l^{(k)})^2\delta_k^2 . 
    \end{split} \label{gan:filter:stability:eq2}
\end{align}
\end{footnotesize}%
\end{subequations}%
Equality (i) expands the inner product of the loss gradients and iterates using $\delta_k$ for the norm of the sum of loss gradients. The inequality (ii) uses the subgradient condition in \eqrefp{gan:filter:subgrad:lemma}, $\gL(\mG_k) - \gL(\mG^{*})\geq  \langle \nabla \gL^{(k)},  \mG_k - \mG^{*}\rangle$ (both for $\gL_a$ and $\gL_u$). 
Rearranging \eqrefp{gan:filter:stability:eq1} and \eqrefp{gan:filter:stability:eq2}, we get 
\begin{small}
\begin{equation}
\begin{split}
 & 2\eta_l^{(k)} \big(\gL_a(\mG_k) - \gL_a^*\big) 
     + 2\eta_l^{(k)} \big(\gL_u(\mG_k) - \gL_u^*\big) \\
     & \leq \E[||\mG_{k} - \mG^{*}||_2^2] - \E[||\mG_{k+1} - \mG^{*}||_2^2] + (\eta_l^{(k)})^2\delta_k^2. \label{gan:filter:iterates:bound1}
\end{split}
\end{equation}
\end{small}%
By summing iterates up to step $K$, we get
\begin{small}
\begin{subequations}
\begin{align}
& 2\Big(\sum_{k=1}^K\eta_l^{(k)}\Big) \min_{k\in[k]} [\gL_a(\mG_k) - \gL_a^*] + \min_{k\in[k]}[\gL_u(\mG_k) - \gL_u^*] \label{gan:filter:iterates:bound:2a}\\
& \explainLeq{iii} 2\sum_{k=1}^K\eta_l^{(k)}[\gL_a(\mG_k) - \gL_a^*] + [\gL_u(\mG_k) - \gL_u^*] \\
& \explainLeq{iv} ||\mG_1 - \mG^{*}||_2^2+\sum_{k=1}^K (\eta_l^{(k)})^2\delta_k^2  \label{gan:filter:iterates:bound:2c}
\end{align}
\end{subequations}
\end{small}%
where (iii) is valid since we take the minimum over all iterations and (iv) is derived from the summation of \eqrefp{gan:filter:iterates:bound1}. Then, arranging \eqrefp{gan:filter:iterates:bound:2a} and \eqrefp{gan:filter:iterates:bound:2c} gives 
\begin{small}
\begin{subequations}
\begin{align}
\begin{split}
& \min_{k\in[k]} [\gL_1(\mG_k) - \gL_1^*] + \min_{k\in[k]}[\gL_2(\mG_k) - \gL_2^*] \\
& \leq \frac{||\mG_1 - \mG^{*}||_2^2 + \sum_{k=1}^K (\eta_l^{(k)})^2\delta_k^2}{2\sum_{k=1}^K\eta_l^{(k)} } \label{gan:filter:iterates:bound:3b}
\end{split}
\end{align}
\end{subequations}
\end{small}%
Thus, if the 2-norm of the vectorized version of $\mG_1 - \mG^*$ is bounded by $r$, 
and with learning rate $\sum_k \eta_l^{(k)} \xrightarrow{} \infty$ but $\sum_k (\eta_l^{(k)})^2 < \infty$, the right hand-side of \eqrefp{gan:filter:iterates:bound:3b} becomes $\frac{r^2 + \sum_k (\eta_l^{(k)})^2 \delta_k^2 }{2\sum_k \eta_l^{(k)}} \xrightarrow{} 0 $. Therefore, using the gradient updates in step1 and step3 minimizes the losses $\gL_a, \gL_u$ and converges to a local optimal point. 

\section{Experiments}
\label{sec:experiments:results}
In this section, we evaluate the capability of our linear filter to (1) generate perturbed smart meter data that reduces the prediction accuracy of sensitive attributes; (2) maintain the minimum energy cost from an optimal control decision using the perturbed data; (3) integrate into a contemporary deep learning architecture with parallelism. The code for our experiments is available at \texttt{\url{https://github.com/markcx/DER_ControlPrivateTimeSeries}}.

\subsection{Setup}


We build up two neural networks to form the adversarial classifier and generator. The adversarial classifier is composed of two fully connected layers with ELU (Exponential Linear Unit) activation to estimate the sensitive attribute from demand. The first layer contains the same number of neurons as the time steps of the meter data series used by the battery optimal controller, and the second layer has half of the neuron numbers of the first layer and outputs a two dimensional vector representing the probability of the associated categories of the label. The generator module is composed of a single linear layer that takes a standard normal random vector and the private labels as inputs, and outputs noise to be added to the original demand. The parameters of the single linear layer form matrix $\mG$. Additionally, we specify $\mG$ to be block diagonal to reduce the number of learning parameters, i.e. $\mG = [\Gamma, \mV]$ where $\Gamma$ is a diagonal matrix. Given the number of columns in our weight matrix is $c_w$ (e.g. the $c_w$ for $\mG$ is $26$ for the solar dataset and $50$ in our residential experiments), we use uniform initialization\cite{he2015delving} between $(-\frac{1}{c_w}, \frac{1}{c_w})$ for both the adversary and generator networks. We use 85\% of the data for training and the remaining 15\% for testing the performance of the filter. We set hyper-parameters $\beta_1=\beta_2=\beta_3 = 10^{-5}, \kappa=10^{-3}$ throughout the experiments. The learning rate for the classifier is $10^{-3}$ and the learning rate for the generator starts from $0.1$ and decays $20\%$ for every 100 steps. We present the classification accuracy to indicate the correlation, as a lower accuracy implies a lower value of mutual information\cite{chen_xiao2019safeml}, thus, there is less correlation between the demand and sensitive labels. We set the initial battery state of charge to 1\% of its maximum energy capacity, i.e. $B_{init} = 0.01B$. We use a time-of-use price structure with two tiers: a high price of \$0.463 per KWh from 4pm-9pm and \$0.202 per KWh for the rest of the day.

\subsection{Examples}
\subsubsection{Integration of storage and solar generation}
\begin{figure}[ht]
    \centering
    \centerline{\includegraphics[width=0.51\columnwidth]{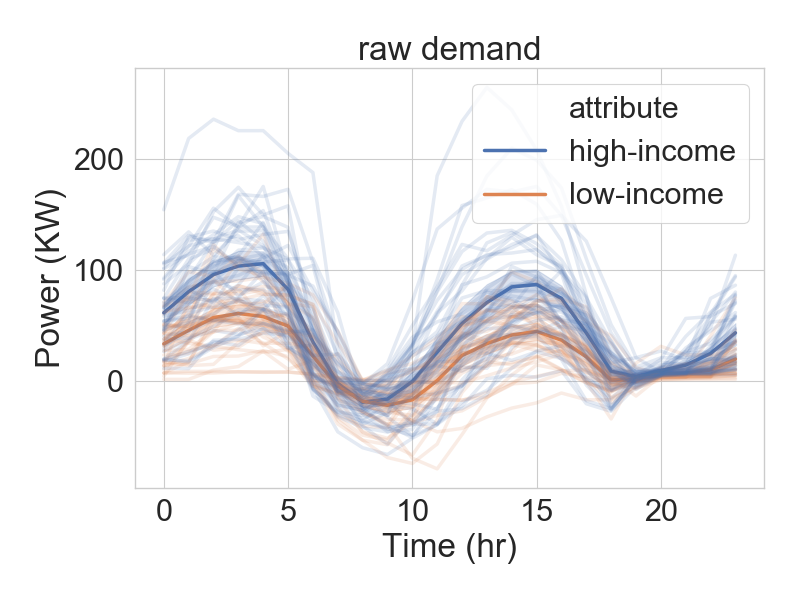}
    \includegraphics[width=0.51\columnwidth]{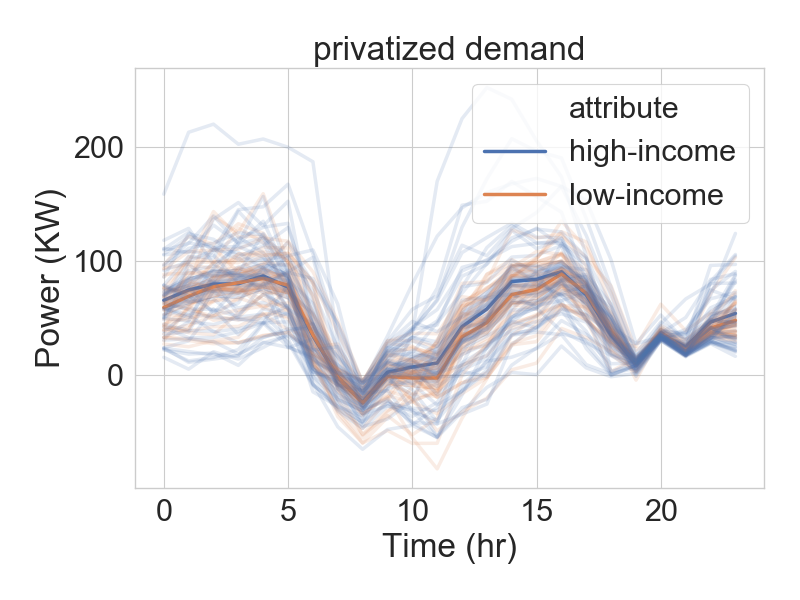}}
    \caption{A batch of 24-hour demand with solar generation that is net negative in certain hours allowing storage to minimize the cost through an optimal charge and discharge sequence. The \textbf{left panel} shows the raw demand. The \textbf{right panel} shows the privatized demand.}
    \label{fig:solar:storage:ctrl}
\end{figure}

\begin{figure}[!ht]
    \centering
    \includegraphics[width=0.85\columnwidth]{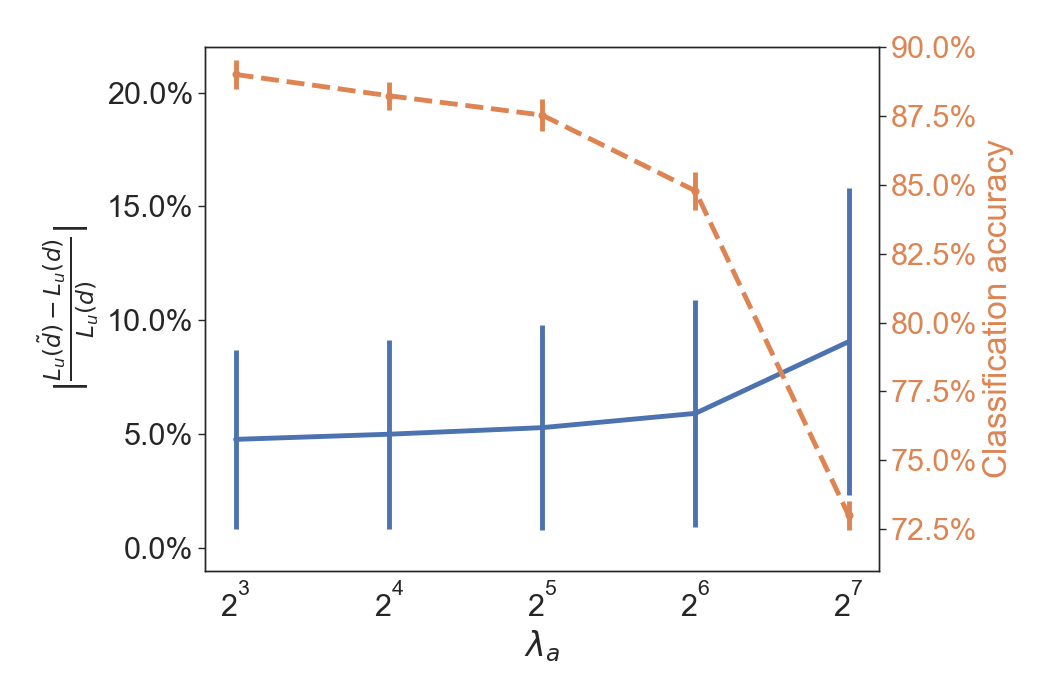}
    \caption{The trade-off between privacy and utility controlled by parameter $\lambda_a$, which places weight on the private attribute classification loss.}
    \label{fig:solar:syn:tradeoff:acc:util:loss}
\end{figure}

For our first experiment, we aggregated 24-hour demand consumption from thousands of homes into groups of 100-200 homes and added solar generation. The aggregations represent the demand seen at a secondary transformer from the perspective of a utility company. The goal is to minimize the energy cost by running the optimal charging and discharging controls for battery storage given a prescribed price. Each demand comes with a binary label indicating if the demand is from a high- or low-income group. We wish to privatize the demand before sending it to the storage operator to perform cost minimization, so the operator cannot infer any sensitive information from its customers. The left panel of \Figref{fig:solar:storage:ctrl} shows the income attribute can be easily inferred from the raw demand as the height of the peaks are clearly distinguishable. The right panel of \Figref{fig:solar:storage:ctrl} shows that the privatized demands are perturbed such that two labels overlap making it harder to tell which demand has high or low income.  
However, there is a trade-off between privacy and utility when perturbing the data. We use the hyper-parameter $\lambda_a$ to balance the adversarial loss and the utility loss i.e. smaller $\lambda_a$ means less weight for privacy and more for utility, as shown in \Figref{fig:solar:syn:tradeoff:acc:util:loss}. When $\lambda_a$ increases from 8 to 128, the classification accuracy of the income label drops from 89.4\% to 73\% as we expected. The raw classification accuracy with zero weight is $95.2\%$. The loss of performance of the cost minimization by using privatized demand instead of raw demand ranges from $5\%$ at $\lambda_a=8$ to almost $10\%$ at $\lambda_a=128$ on average, which shows that high privacy comes with a performance cost for this battery control problem.

\subsubsection{Deployment of storage on residential users} 
The second experiment considers residential customers adopting batteries to minimize their energy cost without selling excess to the grid. The control of the battery is performed by an outside program, so the owner wishes to privatize their demand before sending it to the controller. The dataset is from the Irish CER Smart Metering Project\cite{ucd, beckel2014revealing}. We select a year of meter data for meters that contain a record indicating if they belong to a large or small home and partition it into daily sequences with 48 entries for each day. We end up with 54478 records in total. Recall that our goal is to create altered demand that won't degrade the cost savings while removing the correlation between the demand and the attribute indicating a small or large home. 
%
%
%
%
%
%

\begin{figure}[ht]
    \centering
    \includegraphics[width=0.85\columnwidth]{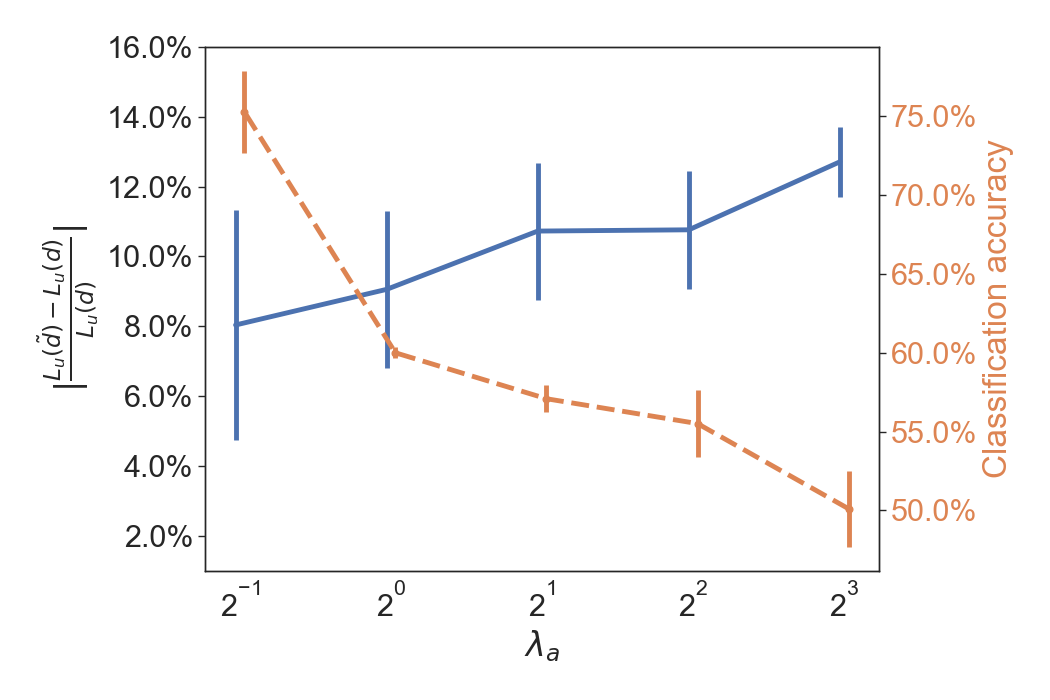}
    \caption{The trade-off between the utility and privacy for the CER dataset\cite{ucd}. The privacy label indicates a large or small home. $\lambda_a$ weighs the privacy loss.} 
    \label{fig:cer:tradeoff:util:priv}
\end{figure}

%
\Figref{fig:cer:tradeoff:util:priv} depicts the trade-off between utility degradation and privacy gain for different weights on privacy loss. The accuracy of classifying large or small homes based on the raw demand is 77.5\%. When we have low weight on the privacy loss (e.g. $\lambda_a=0.5$), the classification accuracy only drops a little to 75\%, with a greater sacrifice on cost saving performance (e.g. increased to 8\% more cost on average). In the high privacy weight scenario, the classification accuracy drops down to 50\% as desired, while the utility performance gap only increases up to 12\%.

\subsection{Parallelism}
The experiments in this section are run on a six-core Intel Core i7 CPU @2.2GHz. Current standard solvers like Gurobi or Mosek without support of in-batch parallelism can be computationally expensive for solving a quadratic problem. Our filter makes use of automatic differentiation for a cone program (DIFFCP)\cite{agrawal2019differentiating} and leverages multiprocessing to speed up the forward and backward calculations.

\begin{figure}[!ht]
    \centering
    \includegraphics[width=0.8\columnwidth]{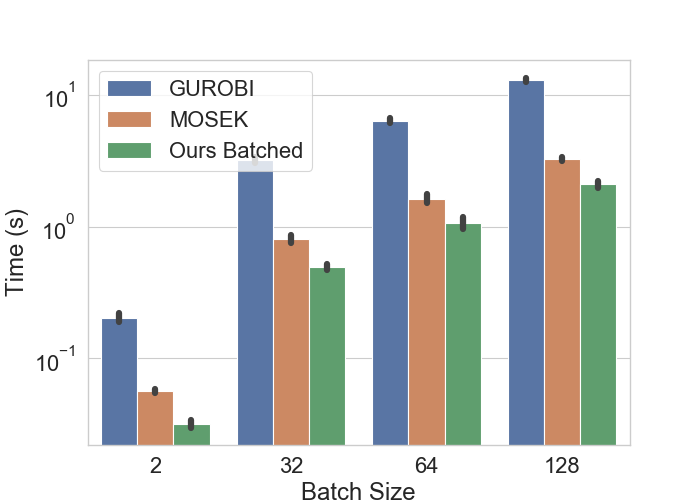}
    \caption{CPU run time of a batched optimization using Gurobi v8.1.0, Mosek v8.1.0.60, and our parallel module. }
    \label{fig:runtime:compare}
\end{figure}

\Figref{fig:runtime:compare} displays the mean and standard deviation of running each trial 8 times, showing that our batched module
outperforms Gurobi or Mosek, which are highly tuned commercial solvers for reasonable batch sizes. For a minibatch size of 128, we solve
all problems in an average of 1.31 seconds, whereas Gurobi takes an average of 11.7 seconds. This speed improvement for a single minibatch makes the difference between a practical and an unusable solver in the context of training a deep learning architecture.
\section{conclusion} \label{sec:conclusion}
We have presented a method for the privatization of personal data that maintains its utility in the optimal control of energy resources. Our method comprises a small linear filter that adds random noise to the data conditional on the private attributes we wish to protect. The linear filter is trained using a minimax optimization procedure that balances the trade-off between classifcation accuracy of the private attributes and the performance of an optimal controller. Additionally, we include a distortion penalty to preserve aspects of the data that are not specified by the utility or privacy functions in order to avoid adding arbitrary noise. We have demonstrated that this method is capable of removing the correlation between the released private data and the sensitive attributes while maintaining limited loss of the utility of the data using two datasets. Limitations of this method include the requirement to solve an optimization in the training loop, which can be computationally intensive for large problems; however, we suspect only a few iterations of the optimization are needed to achieve the desired gradients, which will dramatically reduce the computation required.


\section{Appendix}

\subsection{Battery control details}
\begin{figure}[ht]
    \centering
    \includegraphics[width=0.99\columnwidth]{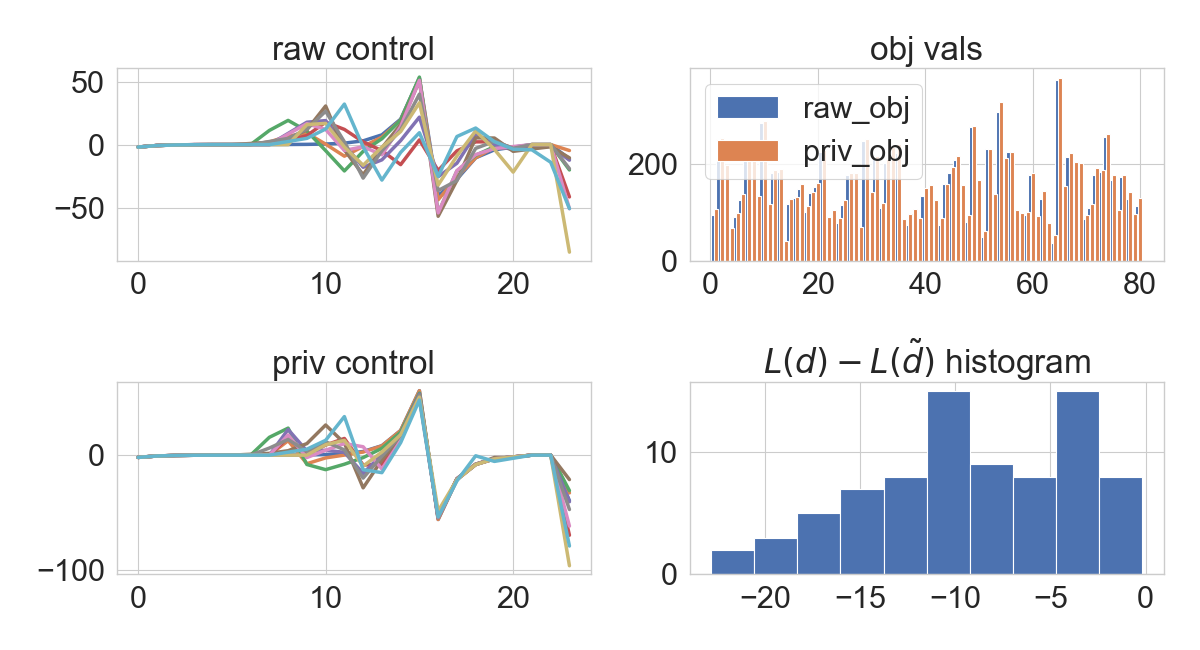}
    \caption{Analysis of storage control for the aggregated homes experiment with $\lambda_a = 128$. The \textbf{upper-} and \textbf{lower-left} panel show the charging and discharging power in kilowatts (KW). Different colored curves represent different samples in the batch. The \textbf{upper-right} panel shows the daily electricity cost when operating the battery using raw or private demand (x-axis is the sample number, y-axis is in dollars (\$)). The \textbf{lower-right} panel shows a histogram of the loss gap. (The x-axis is the increased cost in \$; the y-axis is the number of days that show similar cost increases in a batch. )  }
    \label{fig:syn:ctrl}
\end{figure}

We present a snapshot of the results for the storage control based on the raw and private demand data. \Figref{fig:syn:ctrl} displays the storage control for our experiment with aggregated homes and solar generation. The upper-left and lower-left panel show the 24-hour charging and discharging decisions with each color representing one sample in a batch. The control decisions made with raw versus privatized demand data are closely aligned in general, but have different charging and discharging amounts of power due to perturbation. However, such an altered charging profile doesn't increase the minimum cost much as we can see from the upper-right and lower-right panels of \Figref{fig:syn:ctrl}. The electricity cost increases by a maximum of \$22 USD per day given that the highest daily cost is around US \$390 USD. (Each bin spans the range of \$2.5 USD for \Figref{fig:syn:ctrl}.)      

\begin{figure}[!ht]
    \centering
    \includegraphics[width=0.99\columnwidth]{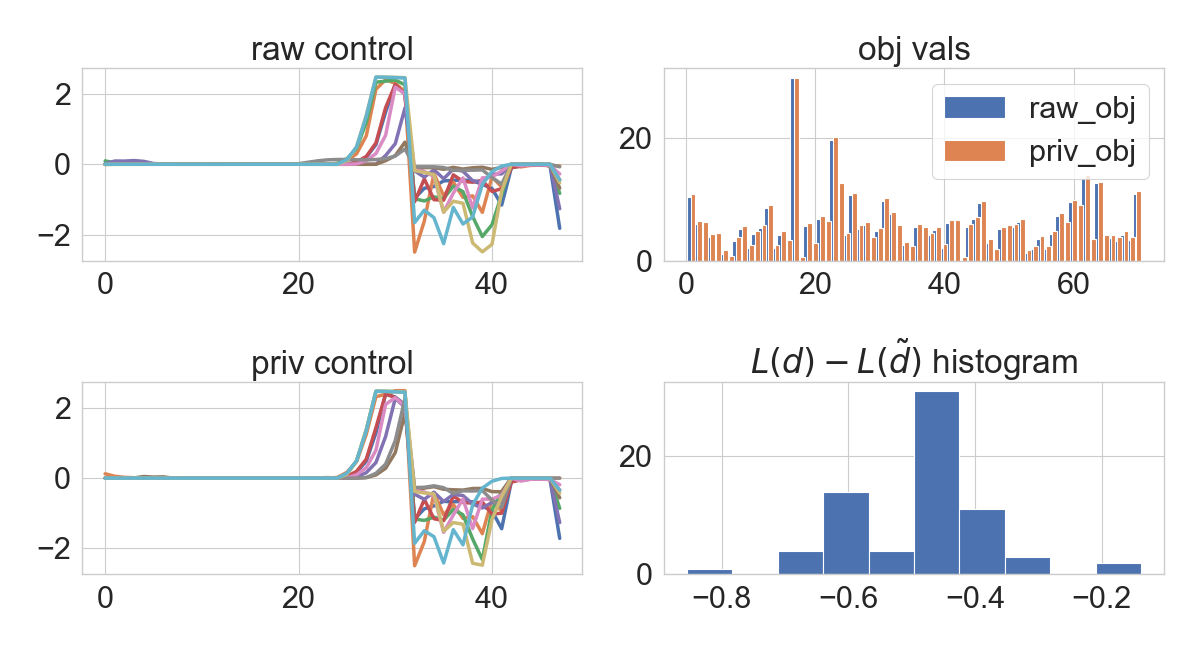}
    \caption{Analysis of storage control for the CER data experiment with $\lambda_a = 8$. Each panel has the same x- and y-axis as \Figref{fig:syn:ctrl}} 
    \label{fig:cer:ctrl}
\end{figure}

\subsection{Quadratic problem}\label{appx:qp:formulate}
A canonical form of the quadratic constrained minimization problem (QP) is expressed as follows: 
\begin{subequations}
\begin{align}
\min_{x} &\quad \frac{1}{2} x^TQx + q^Tx  \label{qp:standard:obj} \\
\text{s.t} & \quad Ax = b  \label{qp:standard:c1}\\
 &\quad Gx \leq h. \label{qp:standard:c2} 
\end{align} 
\end{subequations}

We first show that the basic battery storage problem can be considered as a special case of QP. We start with the 24-hour horizon storage problem in Problem\plaineqref{ctrl:privD:prob1}. 
We can express the constraints from \eqrefp{ctrl:privD:prob1:c3} to \eqrefp{ctrl:privD:prob1:c5} as \begin{footnotesize}
\begin{align}
    \underbrace{\begin{bmatrix}
    I & 0 & 0 \\
    -I & 0 & 0 \\
    0 & I & 0 \\
    0 & -I & 0 \\
    0 & 0 & I  \\
    0 & 0 & -I  \\
    -I & I & 0
    \end{bmatrix}}_{G} \begin{bmatrix}x_{in} \\ x_{out} \\ x_s\end{bmatrix} \leq \begin{bmatrix} 
    c_{in} \\
    0 \\
    c_{out} \\
    0 \\
    B \\
    0 \\
    \vd
    \end{bmatrix} \Leftrightarrow Gx \leq h . \label{qp:constraint:ineqs}
\end{align}
\end{footnotesize}
We add a constraint that the net of the demand and storage is greater than or equal to 0, so we can formulate the objective as a QP. This constraint does not modify the original problem as long as it is feasible because the optimal solution will implicitly make the net of demand and storage greater than or equal to 0. The constraints in \eqrefp{ctrl:privD:prob1:c1}-\eqrefp{ctrl:privD:prob1:c2} are expressed as  
\begin{align}
\begin{split}
    \underbrace{ \begin{bmatrix}
    0   & 0  & 1, \dots 0\\
    [\eta_{in} I , 0] & [-\nicefrac{1}{\eta_{out}}I, 0] & [I, 0] - [0, I] 
    \end{bmatrix}}_{A} \begin{bmatrix}x_{in} \\ x_{out} \\ x_s\end{bmatrix} & =
    \begin{bmatrix} {B}_{init}\\ 0 \end{bmatrix} \\ \Leftrightarrow  Ax & =b, \label{qp:constraint:eqs}
    \end{split}
\end{align}
with $[I, 0] \in \R^{23\times 24}$. The objective \eqrefp{ctrl:privD:prob1:obj} can be converted to a standard QP by letting 
\begin{align}
    \mQ = \begin{bmatrix}
    \beta_1 I & 0 & 0 \\
    0 & \beta_2 I & 0 \\ 
    0 & 0 & \beta_3 I 
    \end{bmatrix}, \quad 
    q = \begin{bmatrix}
    p \\
    -p \\
    -2\beta_3 \alpha B\mathbf{1}
    \end{bmatrix}.
    \label{eq:Q:q:obj:QP:cvx}
\end{align}
Therefore, it is straightforward to discover that $x^TQx + q^Tx$ is the new form of the objective.

%
%
\begin{small}
\bibliographystyle{unsrt}
\bibliography{pscc2020priTSctrl}
\end{small}


\end{document}